\documentclass{article}
\usepackage{amsmath, amssymb, amsfonts}
\title{ Anisotropic fluid from nonlocal tidal effects } 
\author{Hristu Culetu, \\Ovidius University, Dept.of Physics, \\B-dul Mamaia 124, 900527 Constanta, Romania, \\e-mail : hculetu@yahoo.com}

\begin{document}
\numberwithin{equation}{section}
\pagenumbering{arabic}
\maketitle
\newcommand{\fv}{\boldsymbol{f}}
\newcommand{\tv}{\boldsymbol{t}}
\newcommand{\gv}{\boldsymbol{g}}
\newcommand{\OV}{\boldsymbol{O}}
\newcommand{\wv}{\boldsymbol{w}}
\newcommand{\WV}{\boldsymbol{W}}
\newcommand{\NV}{\boldsymbol{N}}
\newcommand{\hv}{\boldsymbol{h}}
\newcommand{\yv}{\boldsymbol{y}}
\newcommand{\RE}{\textrm{Re}}
\newcommand{\IM}{\textrm{Im}}
\newcommand{\rot}{\textrm{rot}}
\newcommand{\dv}{\boldsymbol{d}}
\newcommand{\grad}{\textrm{grad}}
\newcommand{\Tr}{\textrm{Tr}}
\newcommand{\ua}{\uparrow}
\newcommand{\da}{\downarrow}
\newcommand{\ct}{\textrm{const}}
\newcommand{\xv}{\boldsymbol{x}}
\newcommand{\mv}{\boldsymbol{m}}
\newcommand{\rv}{\boldsymbol{r}}
\newcommand{\kv}{\boldsymbol{k}}
\newcommand{\VE}{\boldsymbol{V}}
\newcommand{\sv}{\boldsymbol{s}}
\newcommand{\RV}{\boldsymbol{R}}
\newcommand{\pv}{\boldsymbol{p}}
\newcommand{\PV}{\boldsymbol{P}}
\newcommand{\EV}{\boldsymbol{E}}
\newcommand{\DV}{\boldsymbol{D}}
\newcommand{\BV}{\boldsymbol{B}}
\newcommand{\HV}{\boldsymbol{H}}
\newcommand{\MV}{\boldsymbol{M}}
\newcommand{\be}{\begin{equation}}
\newcommand{\ee}{\end{equation}}
\newcommand{\ba}{\begin{eqnarray}}
\newcommand{\ea}{\end{eqnarray}}
\newcommand{\bq}{\begin{eqnarray*}}
\newcommand{\eq}{\end{eqnarray*}}
\newcommand{\pa}{\partial}
\newcommand{\f}{\frac}
\newcommand{\FV}{\boldsymbol{F}}
\newcommand{\ve}{\boldsymbol{v}}
\newcommand{\AV}{\boldsymbol{A}}
\newcommand{\jv}{\boldsymbol{j}}
\newcommand{\LV}{\boldsymbol{L}}
\newcommand{\SV}{\boldsymbol{S}}
\newcommand{\av}{\boldsymbol{a}}
\newcommand{\qv}{\boldsymbol{q}}
\newcommand{\QV}{\boldsymbol{Q}}
\newcommand{\ev}{\boldsymbol{e}}
\newcommand{\uv}{\boldsymbol{u}}
\newcommand{\KV}{\boldsymbol{K}}
\newcommand{\ro}{\boldsymbol{\rho}}
\newcommand{\si}{\boldsymbol{\sigma}}
\newcommand{\thv}{\boldsymbol{\theta}}
\newcommand{\bv}{\boldsymbol{b}}
\newcommand{\JV}{\boldsymbol{J}}
\newcommand{\nv}{\boldsymbol{n}}
\newcommand{\lv}{\boldsymbol{l}}
\newcommand{\om}{\boldsymbol{\omega}}
\newcommand{\Om}{\boldsymbol{\Omega}}
\newcommand{\Piv}{\boldsymbol{\Pi}}
\newcommand{\UV}{\boldsymbol{U}}
\newcommand{\iv}{\boldsymbol{i}}
\newcommand{\nuv}{\boldsymbol{\nu}}
\newcommand{\muv}{\boldsymbol{\mu}}
\newcommand{\lm}{\boldsymbol{\lambda}}
\newcommand{\Lm}{\boldsymbol{\Lambda}}
\newcommand{\opsi}{\overline{\psi}}
\renewcommand{\tan}{\textrm{tg}}
\renewcommand{\cot}{\textrm{ctg}}
\renewcommand{\sinh}{\textrm{sh}}
\renewcommand{\cosh}{\textrm{ch}}
\renewcommand{\tanh}{\textrm{th}}
\renewcommand{\coth}{\textrm{cth}}

\begin{abstract}
The Shiromizu et al. \cite{SMS} covariant decomposition formalism is used to find out the brane properties rooted from the 5-dimensional Witten bubble spacetime. The non-local tensor $E_{ab}$ generated by the 5-dimensional Weyl tensor gives rise at an anisotropic energy-momentum tensor on the brane with negative energy density and $p = \rho/3$ as equation of state. The tidal acceleration is towards the brane and that is in accordance with the negative energy density on the brane. The anisotropic fluid has vanishing ''bulk'' viscosity but the shear viscosity coefficient is $r$- and $t$- dependent. The brane is endowed with an apparent horizon which is exactly the radial null geodesic.
 \end{abstract}
 
\section{Introduction}

General Relativity is considered to breakdown at very high energies where probably it is a limit of a more general theory. Gravity becomes a truly higher dimensional theory at short ranges, when string theory/M theory applies \cite{RM1,RM2}. 

In the brane-world scenario, standard matter fields are confined to the 4-dimensional spacetime (the brane) whereas gravity propagates in the full spacetime (the bulk) \cite{SSM, SMS}. Shiromizu et al. formulated covariant equations that describe both the 5-dimensional gravity in the bulk and the 4-dimensional gravity on the brane. They found that a positive tension brane has the correct sign of gravity and their equations become the conventional Einstein equations in the low energy limit provided that the nonlocal term rooted from the ''electric'' part of the 5-dimensional Weyl tensor (which describes the tidal forces) is negligible. In addition, there is no \textit{a priori} reason to expect that it is tiny even in the low energy limit \cite{SSM}. 

Maartens \cite{RM2} showed that local effects of the bulk on the brane lead to quadratic corrections of the density, pressure and heat flux. The free gravitational field in the bulk produces nonlocal effects on the brane, including energy density and anisotropic stresses. He also calculated the gravitational (tidal) acceleration of the fluid worldlines, showing how the worldlines have a non-gravitational acceleration off the brane at high energies, which is directed towards the brane. 

Dadhich et al. \cite{DMPR}, Germani and Maartens \cite{GM} and Casadio and Ovalle \cite{CO} (see also \cite{DV} and \cite{HL}) studied models of spherically-symmetric stars and black holes (BH) localized on a three-brane in 5-dimensional gravity in the Randall-Sundrum (RS) scenario. Dadhich et al. have shown that the Reissner-Nordstrom geometry is an exact solution of the effective Einstein equations on the brane, a BH with a tidal ''charge'' arising via gravitational effects from the fifth dimension. The solution satisfies a closed system of equations on the brane, describing a strong-gravity regime. 

Germani and Maartens \cite{GM} proved that the vacuum exterior of a star is not a Schwarzschild spacetime in the RS braneworld model but has stresses induced by 5-dimensional gravitational effects. They also found two different non-Schwarzschild exteriors that matches the star interior on the brane. The ''minimal geometric deformation'' conjecture leads Casadio and Ovalle \cite{CO} to an interior geometry of a spherical star which allows one to map General Relativity to solutions of the effective 4-dimensional brane-world equations with a tidal charge determined by the mass of the source and the brane tension.

The approach we will follow in this paper relies on a well-known vacuum 5-dimensional spacetime \cite{EW} (the so called Witten bubble) and is aiming to study the impact of its gravitational field (expressed by the Riemann/Weyl tensor) upon the three-brane properties. The Witten bubble metric is just an analytical continuation of the 5-dimensional Schwarzschild metric and is known to expand hyperbolically \cite{CJ, BV, OIS, ISO, HC1, HC2}. We shall use the 5-dimensional expanding bubble (which represents the decay of the Kaluza-Klein (KK) vacuum) in the standard spherically-symmetric form. Even though the metric is time-dependent, we found that the Kretschmann scalar and the mixed components of the Riemann tensor are static. Being Ricci-flat, the energy-momentum tensor is vanishing and we have $C^{a}_{~bcd} = R^{a}_{~bcd}$, where $C^{a}_{~bcd}$ and $R^{a}_{~bcd}$ are, respectively, 5-dimensional Weyl and Riemann tensors. By contrast, the kinematical quantities associated to a congruence of timelike worldlines (expansion scalar, shear tensor, etc.) are time dependent. 

By means of the covariant decomposition developed by Shiromizu et al.\cite{SMS} we write down the effective gravitational equations on the brane. The stress tensor $T_{ab}$ on the brane is induced by the 5-dimensional Weyl tensor via the nonlocal tensor $E_{ab}$ from the bulk. We show that the brane metric is conformally-flat (with a Lorentz invariant conformal factor) and the apparent horizon corresponds to a null geodesic.\\
The units are taken such that $c = \hbar = 1$. 

\section{Expanding bubble spacetime}
 Let us consider the analytical continuation of the five-dimensional Euclidean Schwarzschild solution
    \begin{equation}
   ds^{2} = \frac{1}{1 - \frac{R^{2}}{r^{2}}} dr^{2} +  r^{2} (d\chi^{2} + sin^{2}\chi d \Omega^{2}) + (1 - \frac{R^{2}}{r^{2}})dy^{2}
 \label{2.1}
 \end{equation}
where $r \geq R$, $y$ (the fifth coordinate) is periodic of range $2\pi R$ and $d \Omega^{2} = d\theta^{2} + sin^{2}\theta d\phi^{2}$ stands for the metric on the unit 2-sphere. The above metric describes the decay of the standard KK vacuum to a zero energy bubble configuration via a tunneling process. Starting from the initial data $\chi = \pi/2$ slice of (2.1), the Lorentzian evolution of the bubble is obtained through the analytical continuation $\chi \rightarrow igt + \pi/2$ ($g$ is a constant which set the units) \cite{CJ, HW}
 \begin{equation}
  ds^{2} = -g^{2} r^{2} dt^{2} + \frac{1}{1-\frac{R^{2}}{r^{2}}} d r^{2} + r^{2} \text{cosh}^{2} gt~ d\Omega^{2} + (1 - \frac{R^{2}}{r^{2}})dy^{2}
  \label{2.2}
  \end{equation}
which is the Witten bubble solution \cite{EW, HW}. The metric (2.2) is a time-dependent source-free (Ricci flat) solution of the KK field equations, a consequence of a semiclassical decay process of $M^{4} X S^{1}$ which is unstable against a process of semiclassical barrier penetration \cite{BV, BM, HC1, HC2, HC3, HC4}. 

By means of the software package Maple/GRTensor, one obtains for the metric (2.2)\\
- the mixed components of the Riemann tensor
 \begin{equation}
 R^{rt}_{~rt} = R^{r\theta}_{~r\theta} =  R^{r\phi}_{~r\phi} = R^{yt}_{~yt} = -R^{\phi t}_{~\phi t} = -R^{t\theta}_{~t\theta} = -\frac{R^{2}}{r^{4}},~~~R^{ry}_{~ry} = 3\frac{R^{2}}{r^{4}}
  \label{2.3}
  \end{equation}
- the Kretschmann scalar
  \begin{equation}
 R^{abcd} R_{abcd} = \frac{72R^{2}}{r^{8}}
  \label{2.4}
  \end{equation}
- the Weyl tensor 
  \begin{equation}
 C^{a}_{~bcd} = R^{a}_{~bcd}
 \label{2.5} 
  \end{equation}
because the metric is Ricci-flat. Although the geometry (2.2) is not static, we notice that the previous quantities are independent of time. In addition, they are everywhere finite and reach the maximal values at $r_{min} = R$.

Let us take a congruence of ''static'' observers with the velocity vector field $u^{a} = (1/gr, 0, 0, 0, 0),~u^{b} u_{b} = -1$ (the components are written in the order ($t, r, \theta, \phi,y $)). The kinematical quantities associate to the congruence are given by\\
- expansion scalar
 \begin{equation}
 \Theta \equiv \nabla_{b} u^{b} = \frac{2}{r}~ tanh~ gt ,
 \label{2.6}
 \end{equation}
- acceleration 4-vector
 \begin{equation}
  a^{b} = u^{a} \nabla_{a} u^{b} = (0, \frac{1}{r}(1 - \frac{R^{2}}{r^{2}}, 0, 0, 0)
 \label{2.7}
 \end{equation}
with $\sqrt{a^{b}a_{b}} = \frac{1}{r} \sqrt{1 - \frac{R^{2}}{r^{2}}}$.\\ 
- shear tensor
 \begin{equation}
 \sigma_{ab} = \frac{1}{2} (q_{b}^{c} \nabla_{c} u_{a} + q_{a}^{c} \nabla_{c} u_{b}) - \frac{1}{4} \Theta q_{ab} + \frac{1}{2} (a_{b} u_{a} + a_{a} u_{b}) 
 \label{2.8}
 \end{equation}
with
 \begin{equation}
 \sigma^{a}_{~b} = (0,~ -\frac{1}{2r} tanh~gt,~ \frac{1}{2r} tanh~gt, ~\frac{1}{2r} tanh~gt,~- \frac{1}{2r} tanh~gt),
 \label{2.9}
 \end{equation}
where $q_{ab} = g_{ab} + u_{a}u_{b}$ is the projection tensor onto the direction perpendicular to $u_{a}$. In spite of the time-symmetric nature of the spacetime (2.2), the expansion scalar and the shear tensor change sign at $t = 0$ and $\Theta \rightarrow \pm 2/r$ when $t \rightarrow \pm \infty$. Both of them are vanishing when $r \rightarrow \infty$, at constant time.

Let us check whether the Raychaudhuri equation 
 \begin{equation}
 \dot{\Theta} - \nabla_{b} a^{b}+ 2(\sigma^{2}- \omega^{2})+ \frac{1}{4} \Theta^{2} = - R_{ab} u^{a} u^{b}
 \label{2.10}
 \end{equation} 
is fulfilled for the chosen congruence of worldlines. We have above $2\sigma^{2} = \sigma^{ab} \sigma_{ab}$ and $2\omega^{2} = \omega^{ab} \omega_{ab}$. In addition, $ \dot{\Theta} \equiv u^{a} \nabla_{a} \Theta = 2/r^{2} cosh^{2} gt,~\sigma^{ab} \sigma_{ab} = 1/r^{2} tanh^{2} gt, \nabla_{b} a^{b} = 2/r^{2}, \omega_{ab} = 0$ and $R_{ab} = 0$. When these quantities are replaced in (2.10), we conclude that the Raychaudhuri equation is obeyed.

It is well-known the spacetime (2.2) is regular at $r = R$. There is no any event horizon. Being time dependent, we look for an apparent horizon \cite{HC4}. It is obtained from 
 \begin{equation}
 g^{ab}\nabla_{a}P \nabla_{b}P = 0,
 \label{2.11}
 \end{equation} 
where $P(r,t) = rcosh gt$ is the areal radius. Eq. (2.11) yields
 \begin{equation}
 r_{AH}(t) = Rcosh gt.
 \label{2.12}
 \end{equation} 
We see $r_{AH}$ is time dependent, reaches its minimum value at $t = 0$ and is time-symmetric. It is worth noting that (2.12) represents exactly the equation of a free radially moving null particle. Keeping in mind that $r$ is related \cite{HC2} to the Minkowski interval $\bar{r}^{2} - t^{2}$, where $\bar{r} = \sqrt{\bar{x}^{2} + \bar{y}^{2} + \bar{z}^{2}}$, the apparent horizon is given by $\bar{r}_{AH}(\bar{t}) = \bar{t} + R/2$, i.e. the Minkowski light cone. We conclude therefore that the geometry (2.2) is more suitable than Minkowski geometry for the spacetime felt by an inertial observer. The deviation arises only close to the apparent horizon ( near $r = R$ or $t = 0$). In \cite{HC2} (the 2nd paper) we identified the null geodesics $r_{AH}(t)$ with the wormhole throat (see also \cite{HC1}). A similar idea reached Ida et al. \cite{ISO} to whom the brane geometry has the structure of the Einstein-Rosen bridge though they used a different coordinate system.

\section{Brane-world stress tensor}
We take for the time being a general bulk spacetime with five dimensions. Our 4-dimensional world is described by a three-brane embedded in 5-dimensional space. Let $n^{a}$ be the spacelike unit vector field normal to the brane hypersurface and $h_{ab} = g_{ab} - n_{a}n_{b}$, the induced metric on the brane ($g_{ab}$ is the full 5-dimensional metric). Shiromizu et al. \cite{SMS} have shown that, from the Gauss equations relating the Riemann tensors in 5- and 4-dimensions and the Codazzi equations for the variation of the extrinsic curvature, one readily obtains
  \begin{equation}
  \begin{split}
  G_{ab} = (^{5}R_{ab} - \frac{1}{2}g_{ab}~ ^{5}R)h^{a}_{c}h^{b}_{d} + ^{5}R_{cd}n^{c}n^{d}h_{ab} + K_{ab} K - K^{c}_{a}K_{bc}\\ - \frac{1}{2}h_{ab}(K^{2} - K^{cd}_{~cd}) - E_{ab}
  \end{split}
 \label{3.1}
 \end{equation} 
where $^{5}R_{ab}$ is the 5-dimensional Ricci tensor, $K = K^{a}_{~a}$ is the trace of the extrinsic curvature $K_{ab} = h^{c}_{a}h^{d}_{b}\nabla_{c}n_{d}$ and 
 \begin{equation}
 E_{ab} =  ^{5}R^{c}_{~def}n_{c}n^{e}h^{d}_{~a}h^{f}_{~b},
 \label{3.2}
 \end{equation} 
which may be also expressed in terms of the Weyl tensor.

We wish now to apply the Shiromizu et al. formalism for the 5-dimensional metric (2.2). We choose, for convenience, the brane to be located on the hypersurface $y = 0$, so that the normal to the brane is $n^{a} = (0, 0, 0, 0, 1/\sqrt{1 - R^{2}/r^{2}})$. We also have $^{5}R_{ab} = 0$ in (3.1) and $K_{ab} = 0$ because the metric coefficient do not depend on the extra coordinate (it could be checked directly from the definition of $K_{ab}$). It is worth to stress that this is not in contradiction with the Lanczos equations
  \begin{equation}
  K_{ab} - h_{ab}K = - 8\pi G_{5}T_{ab}
 \label{3.3}
 \end{equation} 
($G_{5}$ is the 5-dimensional Newton constant) since the matter energy-momentum tensor $T_{ab}$ on the brane (and the brane tension too) has been chosen to vanish. As Shiromizu et al. \cite{SMS} have noticed, $T_{ab}$ should be evaluated by the variational principle of the 4-dimensional Lagrangean for matter fields, which is missing from the action in our situation. Therefore, (3.1) becomes
  \begin{equation}
  G_{ab} = - E_{ab}
 \label{3.4}
 \end{equation} 
From Eq. (3.2) we get the components of $E^{a}_{b}$
  \begin{equation}
  E^{t}_{~t} = E^{\theta}_{~\theta} = E^{\phi}_{~\phi} = -\frac{R^{2}}{r^{4}},~~~ E^{r}_{~r} = \frac{3R^{2}}{r^{4}}
 \label{3.5}
 \end{equation} 
Now we get the expressions of the stress tensor on the brane via the equations $G_{ab} = 8\pi G_{4} T_{ab}$. Hence \footnote{Strictly speaking, we have to write $G_{ab} = (8\pi G_{5}/L)T_{ab}$, where $L$ is the length scale in the fifth dimension ($R$ in our case). But $G_{5}$ is usually taken as $LG_{4}$, so (3.6) is obtained ($G_{4}$ is the Newton constant in 4 dimensions).}
   \begin{equation}
   8\pi G_{4}T^{a}_{b} = diag\left(\frac{R^{2}}{r^{4}}, -\frac{3R^{2}}{r^{4}}, \frac{R^{2}}{r^{4}}, \frac{R^{2}}{r^{4}}\right). 
 \label{3.6}
 \end{equation} 
The brane geometry (the dimensionally reduced Witten bubble) is obtained taking $y = 0$ in (2.2)
 \begin{equation}
  ds^{2} = -g^{2} r^{2} dt^{2} + \frac{1}{1-\frac{R^{2}}{r^{2}}} d r^{2} + r^{2} \text{cosh}^{2} gt~ d\Omega^{2}.
  \label{3.7}
  \end{equation}
We obtained previously \cite{HC1, HC2, HC3} (see also \cite{OIS, ISO}) the same $T^{a}_{~b}$ but with (3.7) written in its conformally-flat form. 

The energy density of the ''dark'' fluid (3.6) is $8\pi G_{4}\rho = - 8\pi G_{4}T^{t}_{t} = -R^{2}/r^{4}$. It is not surprising that $\rho < 0$ because its origin comes from the free gravitational field in the bulk ( $E^{a}_{~b}$ is rooted from the 5-dimensional Weyl tensor) and, from the point of view of an observer on the brane (which will be considered to be our Universe)  $E^{a}_{b}$ carries influence of nonlocal gravitational degrees of freedom (DOF) from the bulk into the brane, including tidal ''gravito-magnetic'' effects. $E^{a}_{~b}$ is a symmetric tensor, traceless and  $E^{a}_{~b}n^{b} = 0$. We note that it contains bulk DOF which cannot be predicted from data available on the brane. If we take $R$ to be of the order of the 4-dimensional Planck length $l_{P} \approx 10^{-33}cm$, one obtains $\rho \approx - l_{P}^{2}/8\pi G_{4}r^{4} = - \hbar c/8\pi r^{4}$. In other words, $\rho$ has a purely quantum origin ($G_{4}$ no longer appears in its formula) and is similar to the Casimir energy density  between two perfectly conducting parallel plates.

The tidal acceleration in the direction orthogonal to the brane and measured by observers comoving with the fluid from the brane is \cite{RM2}
 \begin{equation}
 A^{y} = -E_{ab}v^{a}v^{b} = -\frac{R^{2}}{r^{4}},
  \label{3.8}
  \end{equation}
where $v^{a} = (1/gr, 0, 0, 0)$ is the 4-velocity of observers comoving with the matter. In other words, $A^{y}$ is towards the brane. Eq. (3.8) shows that localization of gravity near the brane is strengthen by a negative $A^{y}$. That is in accordance with the negative energy density on the brane from nonlocal bulk effects. 

\section{Anisotropic fluid on the brane}
From (3.6) one sees that $T^{a}_{~b}$ corresponds to an anisotropic fluid with $\rho = p_{r}/3 = -p_{\theta} = -p_{\phi}$. If we define a mean pressure
 \begin{equation}
 p = \frac{p_{r} + p_{\theta} + p_{\phi}}{3}
  \label{4.1}
  \end{equation}
one observe that $p = \rho/3$, as for a null fluid. Later we shall prove that $p$ corresponds to the isotropic pressure from the general form of the brane energy-momentum tensor.

The general expression of the brane stress tensor may be covariantly written as 
 \begin{equation}
 T_{ab} = \rho v_{a}v_{b} + pf_{ab} + \pi_{ab} +q_{a}v_{b} +q_{b}v_{a},
  \label{4.2}
  \end{equation}
where $f_{ab} = g_{ab} - n_{a}n_{b} + v_{a}v_{b} = h_{ab} + v_{a}v_{b}$ is the metric felt by comoving observers on the brane, $v_{a}$ is the corresponding 4-velocity , $q_{a}$ is the heat flux and $\pi_{ab}$ is the anisotropic tensor. $T_{ab}$ is diagonal and therefore $q_{a} = 0$. Using (3.6) and (4.2), we get 
 \begin{equation}
 p = -\frac{R^{2}}{3r^{4}},~~~\pi^{r}_{r} = -2 \pi^{\theta}_{\theta}=  -2 \pi^{\phi}_{\phi} = -\frac{8R^{2}}{3r^{4}}
  \label{4.3}
  \end{equation}
The isotropic pressure $p$ is, indeed, given by (4.1) and the anisotropic tensor $\pi_{ab}$ obeys the relations $\pi_{ab}v^{b} = 0,~\pi^{a}_{a} = 0$.

It is known that $\pi^{a}_{~b}$ is related to the viscous properties of the fluid. We therefore consider the expression of $\pi^{a}_{~b}$ in terms of the viscosity coefficients
 \begin{equation}
 \pi^{a}_{~b} = 2\eta \sigma^{a}_{~b} + \zeta \Theta f^{a}_{~b}
  \label{4.4}
  \end{equation}
where here $\sigma^{a}_{~b}$ and $\Theta$ are defined on the brane. $\eta$ and $\zeta$ are, respectively, the shear and ''bulk'' viscosity coefficients. The brane shear tensor is here given by
 \begin{equation}
 \sigma^{r}_{r} = -2 \sigma^{\theta}_{\theta}=  -2 \sigma^{\phi}_{\phi} = -\frac{2}{3r} tanh~gt,
  \label{4.5}
  \end{equation}
with $\sigma^{ab} \sigma_{ab} = (2/3r^{2})tanh~gt$. Using the components of $\pi^{a}_{~b}$ from (4.3), $\sigma^{a}_{~b}$ from (4.5)  and the fact that $\Theta$ has the same expresssion as in (2.6), one finds that 
 \begin{equation}
 \eta(r,t) = \frac{2R^{2}}{r^{3}tanh~gt},~~~\zeta = 0.
  \label{4.6}
  \end{equation}
Than it follows that the fluid has no bulk viscosity but the shear viscosity coefficient is divergent at $t = 0$ where it changes sign and tends to $\pm{2R^{2}/r^{3}}$ when $t \rightarrow \infty$ , at $r = const.$. With all fundamental constants introduced in the expression for $\eta$, we have $\eta(t = \infty) = \eta_{\infty} = (c^{3}/G_{4}) 2R^{2}/r^{3}$. For example, taking as before $R \approx l_{P}$ and $r \approx 1 cm$, one has $\eta_{\infty} \approx 10^{-27} g/cm s$, but $r \approx 10^{-8} cm$ gives $\eta_{\infty} \approx 10^{-5} g/cm s$, much less than water viscosity at $20^{0}C$, which is $\approx 10^{-2} g/cm s$. 

Concerning the Raychaudhuri equation on the brane,  
 \begin{equation}
 \dot{\Theta} - \nabla_{b} a^{b} + \sigma^{ab} \sigma_{ab} - \omega^{ab} \omega_{ab} + \frac{1}{3} \Theta^{2} = - R_{ab} v^{a} v^{b},
 \label{4.7}
 \end{equation} 
one easily verifies that it is observed when we replace in the previous equations $\Theta = (2/r)tanh~gt,~ \dot{\Theta} = 2/r^{2}cosh^{2} gt,~ \nabla_{b} a^{b} = (2/r^{2}) - (R^{2}/r^{4}),~\sigma^{ab} \sigma_{ab} = (2/3r^{2})tanh^{2}gt$ and $R_{ab} v^{a} v^{b} = -R^{2}/r^{4}$.

\section{Conclusions}
The 4-dimensional subspace of the 5-dimensional Witten bubble spacetime is analyzed in this paper. The ground state of the KK theory is unstable against a process of semiclassical barrier penetration. After a short description of the Witten geometry (including the calculation of the curvature invariants, kinematical quantities and the apparent horizon), we took advantage of the Shiromizu et al. formalism to find the effects of Einstein's equations on the brane of constant extra dimension. 

The non-local tensor $E_{ab}$ originating from the 5-dimensional Weyl tensor plays the role of the stress tensor on the brane which represents an anisotropic fluid. The anisotropic stresses give rise to viscous properties  with null bulk viscosity but with a $t$- and $r$- dependent shear viscosity coefficient.

\end{document}